\documentclass[apl,reprint,showpacs,preprintnumbers,amsmath,amssymb, superscriptaddress]{revtex4-1}

\usepackage{graphicx}
\usepackage{bm}
\usepackage{graphicx}
\usepackage{epsfig}
\usepackage{mathrsfs,amssymb}

\usepackage{soul,color}

\include{graphic}

\begin{document}

\preprint{\today}

\title{Quantifying the complex permittivity and permeability of magnetic nanoparticles}

\author{B. M. Yao\footnote{Electronic address:yaobimu@mail.sitp.ac.cn}}
\affiliation{National Laboratory for Infrared Physics, Chinese Academy of Sciences, Shanghai 200083, People's Republic of China}
\affiliation{Department of Physics and Astronomy, University of Manitoba, Winnipeg, Canada R3T 2N2}

\author{Y. S. Gui}
\affiliation{Department of Physics and Astronomy, University of Manitoba, Winnipeg, Canada R3T 2N2}

\author{M. Worden}
\affiliation{Department of Chemistry and Biochemistry, Kent State University, Kent, OH 44242-0001, USA}

\author{T. Hegmann}
\affiliation{Department of Chemistry and Biochemistry, Kent State University, Kent, OH 44242-0001, USA}
\affiliation{Chemical Physics Interdisciplinary Program, Liquid Crystal Institute, Kent State University, Kent, OH 44242-0001, USA}

\author{M. Xing}
\affiliation{Department of Mechanical Engineering, University of Manitoba, Winnipeg, Canada R3T 2N2}

\author{X. S. Chen}
\affiliation{National Laboratory for Infrared Physics, Chinese Academy of Sciences, Shanghai 200083, People's Republic of China}

\author{W. Lu\footnote{Electronic address:luwei@mail.sitp.ac.cn}}
\affiliation{National Laboratory for Infrared Physics, Chinese Academy of Sciences, Shanghai 200083, People's Republic of China}

\author{Y. Wroczynskyj}
\affiliation{Department of Physics and Astronomy, University of Manitoba, Winnipeg, Canada R3T 2N2}

\author{J. van Lierop}
\affiliation{Department of Physics and Astronomy, University of Manitoba, Winnipeg, Canada R3T 2N2}

\author{C.-M. Hu}
\affiliation{Department of Physics and Astronomy, University of Manitoba, Winnipeg, Canada R3T 2N2}


\begin{abstract}
The complex permittivity and permeability of superparamagnetic iron-oxide nanoparticles has been quantified using a circular waveguide assembly with a static magnetic field to align the nanoparticle's magnetization. The high sensitivity of the measurement provides the precise resonant feature of nanoparticles. The complex permeability in the vicinity of ferromagnetic resonance (FMR) is in agreement with the nanoparticle's measured magnetization via conventional magnetometry.  A rigorous and self-consistent measure of complex permittivities and permeabilities of nanoparticles is crucial to ascertain accurately the dielectric behaviour as well as the frequency response of nanoparticle magnetization, necessary ingredients when designing and optimizing magnetic nanoparticles for biomedical applications.
\end{abstract}


\maketitle


Magnetic nanoparticles, with their diameters varying from nanometers to hundreds of nanometers, are under extensive investigation currently for biomedical applications, from MRI contrast agents, microwave imaging enhancement to hyperthermia and drug delivery~\cite{Pankhurst,Berry,Gupta,enancement}.  The principle attraction of magnetic nanoparticles for these applications is that their sizes are significantly smaller than, e.g. target cells, so that magnetic nanoparticles can be in close contact, and their manipulation through their frequency resonant response from external magnetic fields makes it possible to affect or probe cells and tissue.  An outstanding challenge to enable such applications has been a thorough understanding of the nanoparticle's magnetization response (i.e. spin dynamics) when superparamagnetic at high frequencies in complex media such as tissue - knowledge revealed by way of the complex permittivity and permeability that has been problematic to measure.

In principle, a calculation based on the Debye equation and the Kramers-Kr\"onig relations should predict the static and dynamic magnetic behavior of nanoparticles.\cite{Debye,KK1,KK2}  However, values of the required parameters are often an open question as they could be significantly affected by the material, size and measuring device fabrication process.\cite{KK2}  To effectively utilize magnetic nanoparticles for biomedical applications over the frequencies of interest (often in the microwave regime) it is important to accurately characterize not only the magnetic response quantified by the permeability of the nanoparticles, but also the electric response ascertained by the permittivity.\cite{Fluids}

Transmission methods could be used for determining the permittivity ($\epsilon$) and permeability ($\mu$) of material over the requisite wide range of frequencies.\cite{TRL1,TRL2,Fluids}  Usually, $\epsilon$ and $\mu$ are obtained by measuring reflection and transmission coefficients in free space or via waveguide.  Calibration is required to adjust the reference plane to sample surfaces using complicated, multi-step processes including through-reflection-line calibration, which demand expert technique and complex instrumentation.\cite{TRL2}  Irrespective of calibration, an accurate measurement of $\epsilon$ and $\mu$ is very challenging for nanoparticles due to the nanoscale sizes and amounts (typically several mg); the intrinsic nanoparticle scales are usually much smaller than the wavelength range of measurement frequencies.  These challenges can be addressed successfully using cavity perturbation measurement techniques.  In principle, on resonance the response measured via a cavity enables a measure of complex  $\epsilon$ and $\mu$ for the requisite small sample volumes with the necessary precision and accuracy.\cite{RFmethod,Perturbation}  With a nanoparticle sample in a cavity, any changes in the microwave field will result in shifts of the measured resonant frequency and Q-factor.   Unlike the calibration issues surrounding free space- and waveguide-based measurements, the accuracy of cavity-based measurements depends on the frequency resolution of the apparatus and a detailed understanding of the fields in the cavity and sample - issues that are significantly more straightforward to address.

In this Letter, we present a resonant cavity measurement of magnetic nanoparticle's $\epsilon$ and $\mu$ based on a circular waveguide.\cite{Faraday,EPL}  Standing waves are generated inside a circular waveguide cavity (CWC) by involving two transitions (change in cavity shape) at both ends with a 45 degree rotation with respect to the waveguide plane. Based on this setup, a rigorous analytical model is established to provide an accurate calculation of the field distribution.  The use of a network analyzer with a frequency resolution better than one~MHz working over a microwave frequency range allows the for the determination of $\epsilon$ and $\mu$ for very small amounts of nanoparticles ($<$10~mg). The validity and accuracy of these measurements have been verified by mapping the resonant cavity results onto magnetometry measurements on the same nanoparticle systems.


\begin{figure} [t!]
\begin{center}

\includegraphics[width=8.5 cm]{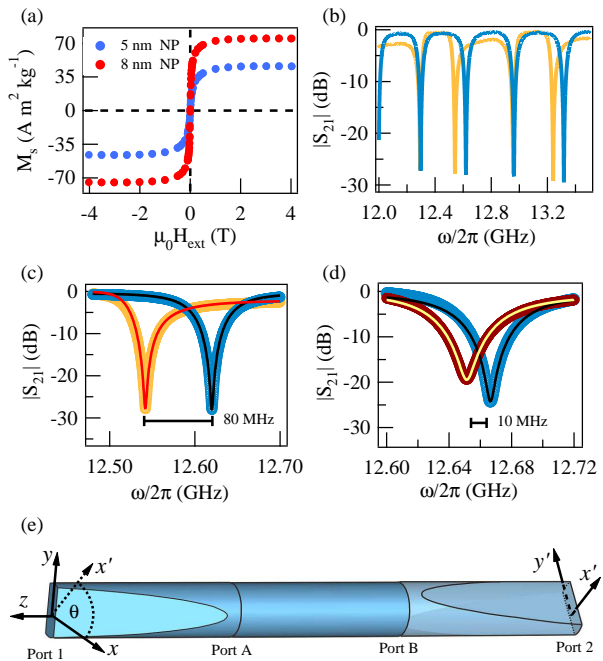}

\caption{(color online) (a) 300~K magnetization vs field measurement of the 5~nm and 8~nm diameter nanoparticles. (b) $S_{21}$ of the empty circular waveguide (blue solid line) and $S_{21}$ with a Teflon sample inserted (yellow solid line).  (c) Lineshape fit for the red-shift caused by Teflon; blue and yellow scatter-points represent the measured $S_{21}$ of the empty waveguide and waveguide with Teflon. Black and red solid lines are the fitted $S_{21}$ from the analytical model.  (d) Lineshape fit for the red shift caused by the iron-oxide nanoparticles; blue and yellow scatter-points represent the measured $S_{21}$ of the empty waveguide and the waveguide with Teflon inserted.  Black and yellow solid lines are the fitted $S_{21}$ from the analytical model. (e)Schematic of the circular waveguide cavity. Two circular-rectangular transitions are rotated by an angle $\theta$ with respect to each other and are connected to a vector network analyzer through coaxial cables.
\label{fig2}}

\end{center}
\end{figure}

Two well characterized nanoparticles were analyzed using the resonant cavity method to determine their $\epsilon$ and $\mu$ values.  One sample was naked iron-oxide nanoparticles with diameters of 5~nm, synthesized using a one-pot technique\cite{Adv}.  The other sample was oleic acid coated iron-oxide nanoparticles, with diameters of 8~nm, made as described in Ref.~\cite{Xing}.  While references~\cite{Adv} and~\cite{Xing} provide detailed characterization of the nanoparticle systems, the salient compositional and microstructural information for these samples is provided in the Supplemental Materials\cite{Supp}, including more detailed field- and temperature-dependent magnetization measurements done using a Quantum Design MPMS XL-5, some of which are shown in Fig.~\ref{fig2} (a).  The nanoparticle systems were clearly superparamagnetic at room temperature, and had saturation magnetization ($M_s$) values of 50$\pm$5~A$\cdot$m$^2$/kg and 75$\pm$5~A$\cdot$m$^2$/kg for the 5~nm diameter and 8~nm diameter nanoparticles, respectively.

The ferromagnetic resonance (FMR) effect in the iron-oxide nanoparticles can be characterized by measuring their broadband transmission coefficients via a coplanar waveguide (CPW).  We have experimentally studied the FMR absorption of the nanoparticles using a CPW, with the measurement details provided the Supplementary Material (SM, Fig.~S2(c)\cite{Supp}).  Qualitatively, the FMR linewidth of the 8~nm sample is larger than that of the 5~nm sample, indicating that the 8 nm nanoparticles with oleic acid coated have a larger damping factor. The measured FMR absorption spectra is normalized by the relative change compared with the background signal. Due to the difficulties in decoupling the background signal from the transmission line, the weak FMR signal of nanoparticles, and the mixing of the waveguide modes, it is problematic to quantitatively determine the complex $\mu$ using CPW measurements.\cite{S21}  For these reasons, we turned to a circular waveguide cavity to measure $\mu$ of the very small amounts of nanoparticles with the necessary high degree of accuracy.


A diagram of the basic experimental setup of the CWC is shown in Fig.~\ref{fig2}(e).  The CWC measurement system has  microwaves transmitted and received with a vector network analyzer (VNA) between ports one and two. Two circular-rectangular transitions are rotated by an angle $\theta$=45 $^{\circ}$ with respect to each other. The wave physics is explained clearly in Refs.~\cite {Faraday} and \cite{Supp}. In Fig.~\ref{fig2} (e), $x$ and $x^\prime$ represent the polarized waves exiting from the circular waveguide to the transitions at ports A or B, and are transmitted without reflection. However, $y$ and $y^\prime$ are polarized waves exiting from the circular waveguide to the transitions at port A and B that are totally reflected with a phase $\phi_{yy}$ and reflection coefficient $R$. Therefore, a standing wave is generated in the long dimension of circular waveguide at specific frequencies.  As shown in Fig.~\ref{fig2}(b), where the blue solid line represents the measured $S_{21}$ parameter for the empty circular waveguide, dips in spectra correspond to frequencies where standing waves are confined in the cavity.  The electric field of the confined standing wave has two spatial distributions. One is that the electric field has a maximum in the mid-plane of the waveguide, which is a symmetric mode.  The other is the electric field with zero amplitude in the mid-plane of the cavity; the anti-symmetric mode.\cite{Faraday}  After a dielectric sample is inserted into the mid-plane position, it only affects the electric field distribution with the maximum in the mid-plane (i.e. the symmetric mode).  As shown in Fig.~\ref{fig2}(b), the yellow solid line represents the measured $S_{21}$ parameter after a 1.6~mm thick Teflon sample is inserted.  The red shift to lower frequency is observed for the symmetric mode, while the resonance frequency remains the same for the anti-symmetric mode.  A detailed description of the  analytic model that describes the wave physics is in the SM\cite{Supp}.  In brief, the measured $S_{21}$ parameter with a sample in the waveguide can be described by a function of permittivity, permeability and frequency, $S_{21}=f(\epsilon ,\mu,\omega)$,  Treating the effective $\epsilon$ and $\mu$ as fitting parameters to model the spectrum, the values can be obtained. 



\begin{figure} [b!]

\begin{center}

\includegraphics[width=8.5 cm]{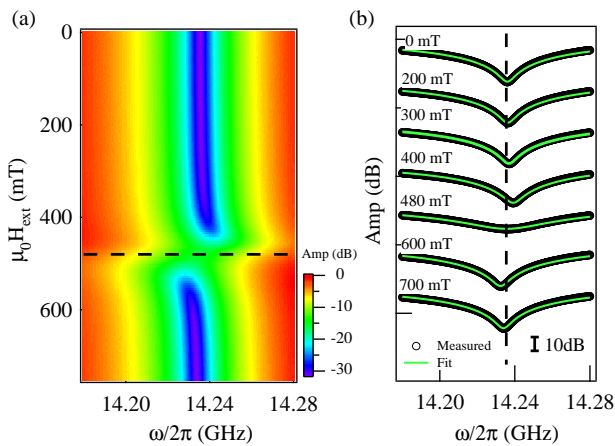}

\caption{(color online) (a) FMR absorption spectra collected by successively changing the magnetic field from 0~mT to 750~mT in an anti-symmetric mode.  The dashed line refers to  the FMR external magnetic field at 14.236~GHz.  (b) Typical FMR absorption spectra where the dashed line refers to the cavity mode without a biasing magnetic field. \label{fig3}}
\end{center}

\end{figure}


Before determining an $\epsilon$ and $\mu$ for a system of interest, the circular waveguide length (24.98~cm), reflected phase, $\phi _{yy}=-101^\circ$, and reflection coefficient $R=0.986$ were obtained by fitting the measured $S_{21}$ of the empty waveguide (Fig.~\ref{fig2}(c)).  After inserting a 1.6~mm thick Teflon sample in the mid-plane of the waveguide, fitting the resultant lineshape established that the real and imaginary components of the Teflon's permittivity to be $\Re(\epsilon$)=2.07 and $\Im(\epsilon$)=0.0008, in good agreement with the reported values\cite{Teflon}. 

With the accuracy of CWC-based system established, $\epsilon$ for 28~mg of the 5~nm iron-oxide nanoparticles was measured by sandwiching the powder of nanoparticles between two layers of transparent adhesive (Scotch, 3M\textregistered) tape and distributed uniformly in the cross-sectional plane and fixed firmly in the mid-plane of the waveguide. $S_{21}$ of the empty circular waveguide with only two layers of tape was measured for the background signal.  The nanoparticle sample presented a clearly observable redshift in the FMR response around 15~MHz, shown in Fig.~\ref{fig2}(d).  To determine the nanoparticle's $\epsilon$ requires knowing the sample thickness. With the particles pressed into a compact cylinder, the effective thickness $l_s$ was estimated by $l_s=m/(A\rho)$, with $m$ the sample mass, $A$ the cross-sectional area of the waveguide, and $\rho$ the density of iron-oxide ($\sim$5.17~kg~m$^{-3}$ from estimates of the electron density of the iron-oxide from x-ray diffraction pattern Rietveld refinements\cite{Supp}). Due to the inevitable presence of air in the nanoparticles sample, the effective thickness $l_s$ is used instead of the physical thickness of nanoparticles sample to avoid the error in the determination of effective parameters $\epsilon$. In addition, $\epsilon$ is determined for frequencies above 12~GHz, where with no applied external magnetic field, the working frequency range is far from the natural resonance frequency of the nanoparticles (obtained from the intercept of $\omega-\mu_0H$ relation in the SM,Fig.~S2(c)) of 1.06~GHz~\cite{Supp}.  Hence, the $\mu$ of the iron-oxide was assumed to be one. As shown in Fig.~\ref{fig2}(d), fitting the lineshape identifies $\Re(\epsilon$)=11.6 and $\Im(\epsilon$)=1.4 for the 5~nm iron-oxide nanoparticles.


\begin{figure} [b!]

\begin{center}

\includegraphics[width=8.5 cm]{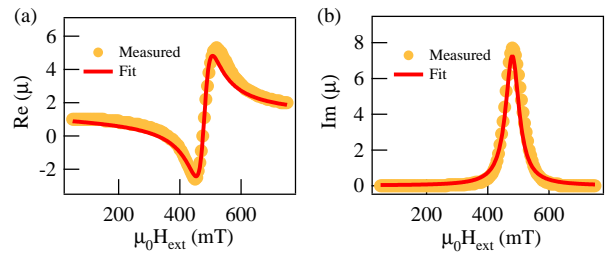}

\caption{(color online) Measured real part (a) and imaginary part (b) of the permeability of the 5~nm diameter iron-oxide nanoparticles. \label{fig4}}
\end{center}

\end{figure}

Determining the $\mu$ of the 5~nm iron-oxide nanoparticles was performed with a varying applied magnetic field ($\mu_0H_{ext}$). For the purpose of enhancing the signal from the sample, the sample-to-cavity volume ratio was increased by shortening the length of circular waveguide by 15.45~cm (from 24.98 cm to 9.53 cm). Moreover, since the measurement to determine $\mu$ relies on the anti-symmetric mode (the electric field is zero) the magnetic field plays the dominant role.  By applying a biasing field between 0 to 750 mT (shown in Fig.~\ref{fig3}(a)) the circular waveguide resonance mode under each magnetic field has been successively collected, and is presented in a compact map with the colour representing the $S_{21}$ amplitude. Clearly, the resonance dip, coloured in blue, indicates a blue shift to higher frequencies with the field going from 0 to 430~mT, which can be explained by the $\Re(\mu)$ being smaller than zero.  Notice that the resonance dip quickly changes from blue-shifted to red-shifted when the field increases from 430 to 530~mT as the FMR resonance frequency (Fig.~\ref{fig3}(a) with dashed lines) approaching the cavity mode value.  With fields above 530~mT, the FMR resonance frequency is higher than the cavity mode, and the observed resonance dip in the mapping gradually goes back to the cavity mode value. Besides the frequency shift, the signal amplitude was found to have significant absorption between 430 and 530~mT due to FMR resonance.

For the purpose of more clearly observing the resonance lineshapes, the field dependence of typical $S_{21}$ parameters are presented in Fig.~\ref{fig3}(b) with a vertical offset.  The dashed line represents the $\mu_0H_{ext}$=0 cavity mode. The absorption at resonance with changing $\mu_0H_{ext}$ can be clearly observed. A 10~dB resonance amplitude difference, due to the strong absorption from the FMR effect is evident at 480~mT when compared to the  0~mT signal.  By taking the complex permeability as fitting parameter, lineshapes for various $\mu_0H_{ext}$ are well described with our analytical model\cite{Supp}.  The resulting $\Re(\mu$) and $\Im(\mu$) are shown in Figs.~\ref{fig4}(a) and (b), respectively.  The modelled $\mu(\mu_0H_{ext})$ relation is also shown, and with the resulting damping factor $\alpha$ and saturation magnetization $M_s$ from the fits.  $\alpha=0.053$ and $M_s=59$~A$\cdot$m$^2$/kg was determined, in excellent agreement with the magnetization measurements performed on the sample nanoparticles ($M_s$=50$\pm$5~A$\cdot$m$^2$/kg, Fig.~\ref{fig2}(a)).  A 180$^{\circ}$ phase delay with $\mu_0H_{ext}$ was determined from both measured $S_{21}$ and fitted $\mu$, with the phase delay a typical feature of resonance, indicating clearly that $\mu$ is the resonant permeability from the ferromagnetic resonance effect.


\begin{figure} [t!]

\begin{center}

\includegraphics[width=8.5 cm]{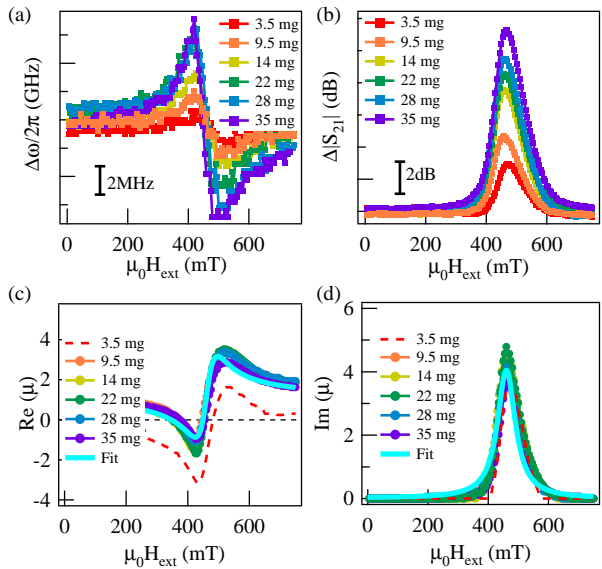}

\caption{(color online) (a)Frequency shift and (b) resonance absorption for different amounts of nanoparticles. (d)Real permeability and (e) imaginary permeability of the various nanoparticle amounts. \label{fig5}}
\end{center}

\end{figure}

To confirm the reliability of the measurements and analysis, 8~nm nanoparticles were also examined.  Additionally, since the measured $\mu$ should be independent of the amount of sample in the cavity, $\Re(\mu$) and $\Im(\mu$) was determined for 3.5 to 35~mg samples of these nanoparticles.  The results for the range of $\mu_0H_{ext}$ are presented in Figs.~\ref{fig5}(a) and (b) respectively. As expected, the frequency shift and absorption amplitudes depend on the amount of sample in the cavity, with more sample providing greater frequency shifts and absorption amplitudes. The nanoparticles are coated with oleic acid which accounts for $\sim$12\% of the total sample mass (from thermogravimetric analysis\cite{Xing}). By using both $\Re(\mu)$ and $\Im(\mu)$ as fit parameters, the complex permeability has been determined for each sample mass, shown in Figs.~\ref{fig5}(c) and (d). While the amount of sample in the cavity increases by an order of magnitude, the variation in $\mu$ is $\pm$10\%.  Fig.~\ref{fig5}(c) shows that the $\Re(\mu)$ of 3.5~mg sample (red dashed line) presents a downward shift compared with $\Re(\mu)$ of the higher mass samples. This is due simply to the observed frequency shift of this sample being close to the measuring precision of VNA ($\sim$0.5~MHz). Nevertheless, the values of real and imaginary components of the permeability are still of a reasonable order. With $\alpha$ and $M_s$ as fitting parameters to describe permeability from the measurements, a fitted $\alpha$ of 0.072 resulted, and $M_s=71$~A$\cdot$m$^2$/kg, in good agreement with the magnetometry $M_s$=75$\pm$5~A$\cdot$m$^2$/kg (Fig.~\ref{fig2}(a)).

In summary, a circular waveguide cavity was used to perform a high precision measurement of the complex permittivity and permeability of tiny amounts of nanoparticles. Combined with a biasing external magnetic field, the resonant permeability under ferromagnetic resonance was measured, and the reliability of the results and analysis was supported by conventional magnetometry.  Our results paves the way for the characterization of the necessarily small amounts of nanoparticles with magnetic resonance, and provides crucial information for developing the optimal uses of magnetic nanoparticles for applications in biomedicine. 
    

We would like to thank Sandeep Kaur for useful discussions.  This work has been funded by NSERC, CFI, CMC and University of Manitoba URGP grants.  BMY was supported by the National Key Basic Research Program of China (2011CB925604), National Natural Science Foundation of China Grant No. 11429401 and STCSM (14JC1406600). TH acknowledges financial support from the Ohio Third Frontier Program for Ohio Research Scholars.

\end{document}